# Title: Kagome modes as a new route to ultra-low thermal conductivity

**Authors:** D.J. Voneshen[1]*, M. Ciomaga Hatnean[2], T.G. Perring[1], H.C. Walker[1], K. Refson[3,1], G. Balakrishnan[2] and J.P. Goff[3].

**Affiliations:**

[1]ISIS Pulsed Neutron and Muon Source, Rutherford Appleton Laboratory, Chilton, Didcot, OX11 0QX, UK.

[2]Department of Physics, University of Warwick, Coventry, CV4 7AL, UK

[3]Department of Physics, Royal Holloway, University of London, Egham, TW20 0EX, UK

*Correspondence to: david.voneshen@stfc.ac.uk.

**Abstract:** From next generation gas turbines to scavenging waste heat from car exhausts, finding new materials with ultra-low thermal conductivity ($\kappa$) has the potential to lead to large gains in device efficiency. Crystal structures with inherently low $\kappa$ are consequently desirable, but candidate materials are rare and often difficult to make. Using first principles calculations and inelastic neutron scattering we have studied the pyrochlore $La_2Zr_2O_7$ which has been proposed as a next generation thermal barrier. We find that there is a highly anharmonic, approximately flat, vibrational mode associated with the kagome planes. Our results suggest that this mode is responsible for the low thermal conductivity observed in the pyrochlores and that kagome compounds will be a fruitful place to search for other low $\kappa$ materials.

**Introduction:** Managing heat flow is vital to the efficient operation of modern technology. For example, high performance thermoelectric devices, which can turn waste heat back into useful power, requires the reduction of heat flow across the device (*1*). Meanwhile, jet turbines now routinely operate with temperatures in the combustion chamber well in excess of the melting temperatures of the superalloy blades (*2*) to improve fuel efficiency (*3*). To prevent the blades melting the heat load is managed by active cooling (in the form of cooling air passed through the blades) and a thermal barrier coating (TBC) applied to the surface (*4*).

The conditions a TBC is subject to places extreme demands on the material properties. A low thermal conductivity ($\kappa$) is key, however, they must also be tough, match the thermal expansion of the blade and resist oxidation (*4*). To push engines to higher operating temperatures requires a material which can meet these challenging demands while having a lower thermal conductivity than the widely used $ZrO_2$ doped with 7 wt% $Y_2O_3$ (7YSZ). One possible candidate is the pyrochlore zirconate $La_2Zr_2O_7$ (LZO) which has a lower thermal conductivity than 7YSZ (*5*) and unlike 7YSZ has no structural phase transitions until it melts at 2300 °C (*6*).

There are several possible origins of this extremely low $\kappa$ in LZO (and the pyrochlores more generally (*7-9*)) but there exists no clear consensus. The pyrochlore lattice is known to develop oxygen vacancies (*10*) and have rare-earth – transition-metal site mixing (also known as stuffing) (*11*). Stuffing in the zirconate pyrochlores can occur for all members (although it is least prevalent in LZO) (*12*) and either defect could give rise to phonon scattering. It has also been suggested that deliberate doping on the rare earth site with a smaller ion can, by analogy to the rattling ions found in other low $\kappa$ materials (*13*), give rise to strong phonon-phonon scattering (*14*). It has even been proposed that these rattling modes exist in the undoped case (*15*). The pyrochlore lattice has generated huge interest as a model system to study geometric magnetic frustration in 3D via corner-sharing tetrahedra (*16*) and in 2D through the kagome planes (*17*). Interestingly, it has been speculated that the kagome planes in pyrochlores may support unusual phonon modes (*18*).

Inelastic Neutron Scattering (INS) offers the opportunity to measure the phonon dispersion across the whole Brillouin Zone and investigate these proposals. Here, we report the first measurements of the phonon dispersion in LZO supported by Density Functional Theory (DFT) calculations. We are able to identify a new class of strongly anharmonic phonon modes associated with the rare earth kagome lattice (shown in figure 1) which are analogous to rather unusual phonon modes found in ball and spring models based on the kagome lattice (*19*) (Fig. 1c). We propose that these kagome modes are the origin of the ultra-low thermal conductivity in the pyrochlores.

**Results:** INS measurements were performed on two co-aligned single crystals of LZO at 5 and 300 K using the MERLIN spectrometer at ISIS (*20*). DFT calculations were performed using the CASTEP code (*21*). A typical cut through the data and corresponding calculation is shown in figure 2. To obtain the strong agreement between calculation and experiment across multiple Brillouin zones, DFT must accurately describe the real space atomic motion. A comparison of the two dispersions along a path through the high symmetry points is also important and shown in figure 3. Here again the agreement is relatively good with the notable exception of the lowest energy mode around the L point in reciprocal space. Experimentally it is almost completely flat but in the calculations it shows a pronounced softening. This mode consists almost entirely of motion on the La sites and at the zone centre the real space atomic motion is shown in figure 1b.

This mode is clearly unusual and a frozen phonon calculation at $\Gamma$ shows it to be strongly anharmonic (see Fig 4a). This is important as it is the anharmonic terms that allow for three-phonon and higher processes which reduce the phonon lifetimes and lower $\kappa$. It is also at relatively low energies (experimentally 8.6 meV) and as figure 3a shows rather flat. This immediately draws comparisons to the rare earth rattlers proposed in the skutterudites (*13*). It is important to note however that while previous computational work has suggested this mode may be a rattler (*15*) it is not. A rattling mode is an incoherent motion of the guest atom within a cage-like crystal structure while here, the kagome plane moves coherently and thus it is a completely different physical phenomenon.

DFT suggests rattlers in the skutterudite $LaFe_4Sb_{12}$ (where La is the rattler) would be found at 9.2 meV and be weakly anharmonic (*22*). Our DFT calculations of the LZO kagome mode show it at a similar energy to these rattlers but the anharmonicity appears to be significantly stronger. We can attempt to quantify this by considering the ratio of quartic to quadratic terms in a calculation of the potential energy landscape obtained from a frozen phonon calculation. In this way, a more strongly anharmonic system will have a larger number. For the La skutterudites this ratio is 0.22 (*22*) while the kagome mode is 3.80.

Experimentally, anharmonicity giving rise to phonon-phonon scattering will appear as a temperature dependent energy broadening of the peak. This is shown for the kagome mode in figure 4b at low and high temperature. The 5K data is consistent with being resolution limited (see methods). In order to fit the 300 K data, the background-subtracted low temperature data was convolved with a Lorentzian of width Gamma = 0.269(7) meV. The absence of evidence for broadening at low temperature suggests that disorder is not the primary mechanism for phonon scattering. The clear evidence for broadening at 300 K shows that the kagome mode meets the requirements for a mode to scatter phonons and lower the thermal conductivity. It however raises an interesting question. Why is this coherent mode in a system without proximity to a phase transition so strongly anharmonic?

**Discussion:** The phonon dispersion from an isolated kagome plane has been studied theoretically with ball and spring models (*19*). If a model is constructed where each site is connected by an identical spring to its four nearest neighbors then it is possible to collapse the lattice for zero energy cost by counter-rotating linked triangles (as shown in figure 1c). This is known as a floppy mode (*23*), a special class of rigid-unit-modes (*24*) (which have themselves attracted much interest in the field of negative thermal expansion (*25*)). Experimentally these floppy modes have been studied in 3D (for example in amorphous silicates (*26*)) and floppy modes in the kagome lattice have been studied in two-dimensional Silica (*27*). In the 2D case, the floppy mode is displaced from zero energy due to next-nearest-neighbor perturbations. Calculations found it to be strongly anharmonic and when pressure was applied became imaginary (implying a structural distortion is favorable). This energy landscape and pressure dependence arise from the mode's nature as an avenue of collapse for the lattice.

In LZO (and the pyrochlores generally) the opposite pressure dependence has been found in modes which we now identify as kagome modes (*15,28*). As the lattice is expanded in DFT calculations the kagome mode softens around the *L* point, eventually becoming imaginary. This is somewhat surprising as we might expect if it is a true analogy to the floppy modes in the kagome lattice then it would have the same pressure dependence. However, in the pyrochlore lattice the kagome planes are filled by the transition metal which provides a much stronger perturbation than the next-nearest neighbor interactions in 2D Silca. It is reasonable to expect that the filling would make any distortion of the kagome planes energetically unfavorable unless the lattice was expanded to reduce $La^{4+}$-$Zr^{3+}$ repulsion.

A recent publication (*29*) has suggested that a number of pyrochlores including LZO exhibit static structural distortions with ionic displacements resembling β-cristobalite (*30*). We note that the disorder in β-cristobalite is dynamic, and we suggest in that case the dynamics of the buckled kagome planes resemble those described here for the kagome planes in LZO. We note further that dynamic β-cristobalite undergoes a structural phase transition to static α-crystobalite, and that no such structural phase transition is observed in pyrochlores. The geometry optimisation of our DFT calculation for LZO finds no evidence for static structural distortions, and recent diffuse neutron scattering studies of pyrochlores confirm that the disorder in the purest compounds is dynamic (*31*). Hence, we are instead able to account for this behaviour from the lattice dynamics.

The presence of highly anharmonic kagome modes in LZO seems very likely to be responsible for its ultra-low κ. Evidence for temperature dependent phonon-phonon scattering of these modes further supports this picture. Such a result has wide implications. There exist a huge range of kagome lattice containing compounds which could be considered for low κ applications. Furthermore, a number are metallic and some recent work has found that compounds with kagome lattice structures are good thermoelectrics (*32*). We propose that the kagome lattice is not just a fascinating playground for magnetism, but also a source of unusual, technologically relevant thermal physics.

**References:**
1. Snyder, G.J. & Toberer E.S. Complex thermoelectric materials. *Nat. Mater*, **7**, 105-114 (2008).

**Acknowledgments:** DJV wishes to acknowledge the contribution of Rebecca Fair from STFC Scientific Computing who optimized the code used to produce figures 2b and 3b.

**Funding:** Experiments at the ISIS Neutron and Muon Source were supported by a beamtime allocation (RB1710105) from the Science and Technology Facilities Council. Computing resources were provided by STFC Scientific Computing Department's SCARF cluster.

**Author contributions:** DJV and JPG conceived the project. DJV devised the experiment and wrote the beamtime proposal with input from JPG, GB and MCH; MCH and GB grew the single crystals. DJV and HCW performed the MERLIN experiment. DJV reduced and analyzed the MERLIN data. DJV performed the CASTEP calculations with advice from KR. TGP performed the resolution fitting using data supplied by DJV. KR made the key connection between the kagome mode and rigid unit modes. DJV wrote the paper and all authors were involved in the review and editing.

**Competing interests:** Authors declare no competing interests.

**Data and materials availability:** The unprocessed neutron data are curated by ISIS can be accessed from doi:10.5286/ISIS.E.86387462. The scripts used to produce the data shown within the paper (both experimental and computational) are available from the corresponding author on reasonable request.

**Supplementary Materials:**

**Materials and Methods**
Single crystals of $La_2Zr_2O_7$ were grown by the floating-zone technique using a four-mirror Xenon arc-lamp optical image furnace (CSI FZ-T-12000-X_VI-VP, Crystal Systems, Inc., Japan) (*33,34*). The growths were carried out in air at ambient pressure and at growth speeds in the range 10 – 15 mm/h. The two rods (feed and seed) were counter-rotated at a rate of 20-30 rpm. The crystal quality was checked using a Laue X-ray imaging system with a Photonic-Science camera.

Two single crystals of total mass 3.639 g were attached to aluminium sample mounts with aluminium wire and aligned in the *HHL* scattering plane. Inelastic neutron scattering measurements were made on the MERLIN spectrometer (*20*) at the ISIS pulsed neutron and muon source. MERLIN was set up with the Gadolinium chopper running at 300 Hz and focused on 35 meV. The sample was loaded into the standard MERLIN closed cycle cryostat and measurements were made at 300 and 5 K. The sample was rotated about its vertical axis over the angular range -165 to -87° in 1° steps where 0° is defined as the 00*L* direction being parallel to the incident beam. The data was reduced with Mantid (*35*) and combined, symmetrised and analysed with Horace (*36*).

Calculations of the phonon dispersion were performed with DFT as implemented in the CASTEP code (*21*). Default ultrasoft pseudopotentials (version 17.2.1) were used and the results presented used the Local Density Approximation (LDA) to exchange and correlation. A plane wave cut-off of 600 eV with electronic sampling on a Monkhorst-Pack mesh of 4 by 4 by 4 within the primitive cell was found to reduce the error in the forces to below 0.005 eV Å$^{-1}$. The structure and lattice parameters were relaxed with the quasi-Newton method (*37*) corrected for the finite basis set (*38*). Phonon calculations were performed using the finite displacement/supercell method (*39*) within a single cubic unit cell.

The instrumental resolution function was computed using TOBYFIT (40) convolved with the dispersion calculated by CASTEP for the kagome mode with a constant energy offset to bring the dispersion in line with the measured one. It was assumed that the intensity was constant across the integration range ($3.25 \leq h \leq 4.75$, $5.25 \leq k \leq 5.75$ and $6.25 \leq l \leq 6.75$). In addition to the scattering by phonons, there is scattering from the cryostat, sample holder and incoherent scattering from the sample (which although small is non-zero). Furthermore, at 300 K the broadening of the excitations causes them to overlap making individual treatment impossible. A measurement was made of the cryostat with the empty sample holder (sample was untied and the same Al wire reattached to the mount so the exact same mass of Al was in the beam) at both 5 and 300 K and this was used to perform a background subtraction.

At 5 K there is no overlap between the kagome mode and other nearby modes. The width of Lorentzian required to fit the 5 K data is strongly correlated with the value of the flat background and this is reflected in large error bars similar to the magnitude of the width (the best fit was obtained with $\varGamma = 0.26\ (27)$ meV). As the data is consistent with being resolution limited it used to fit the high temperature data. The background subtracted low temperature data was used as the resolution function and convolved with a Lorentzian, to give the value quoted in the main text. After correction for Bose-Einstein statistics the only free parameter to the fit was the width of the Lorentzian.

References (*33-40*)

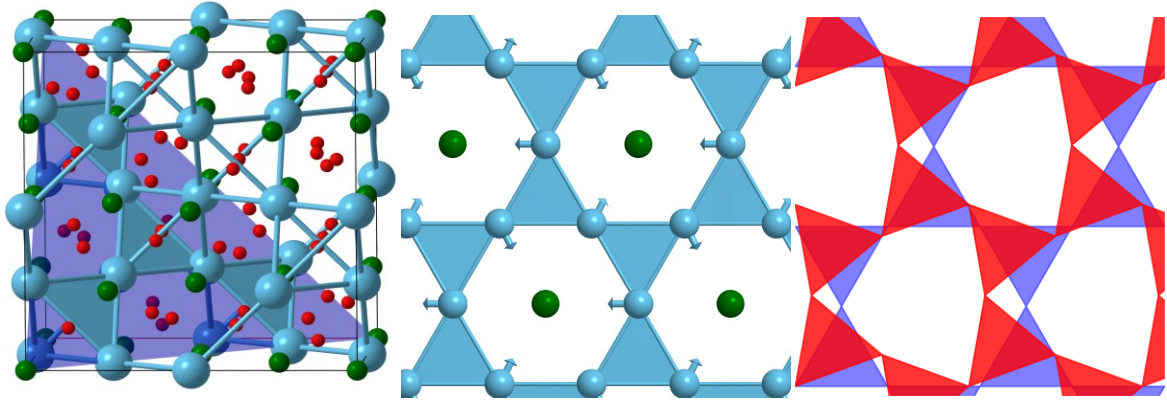

**Fig. 1.** Crystal structure and relation to kagome vibrations (a) The conventional cubic cell of the pyrochlore $La_2Zr_2O_7$ with La in light blue, Zr in green and O in red. A kagome plane normal to the [111] direction is highlighted in blue. (b) A kagome mode in LZO with the ionic displacements shown. There is almost no motion of the Zr, O or the La site linking (111) planes together. (c) An unfilled kagome plane has collapse modes known as floppy modes (*19*) shown here in red. The red lattice has been obtained by rotating the bottom left triangle by 20° and fixing the size of the triangles.

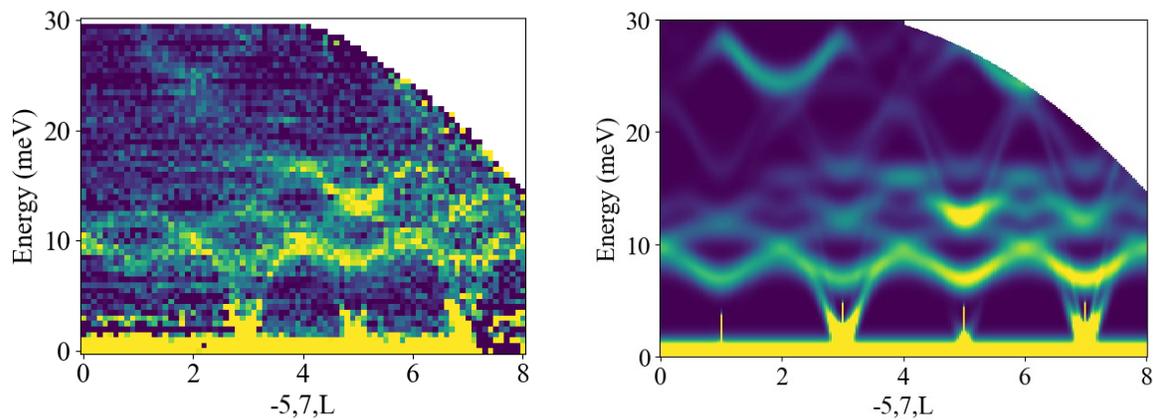

**Fig. 2.** Experimental and calculated phonon dispersion along [-5,7,*L*] (a) A typical cut through the INS data from LZO at 300 K. (b) The calculated one phonon scattering for this cut from DFT. Both plots use a linear color scale with an arbitrary scaling between the two.

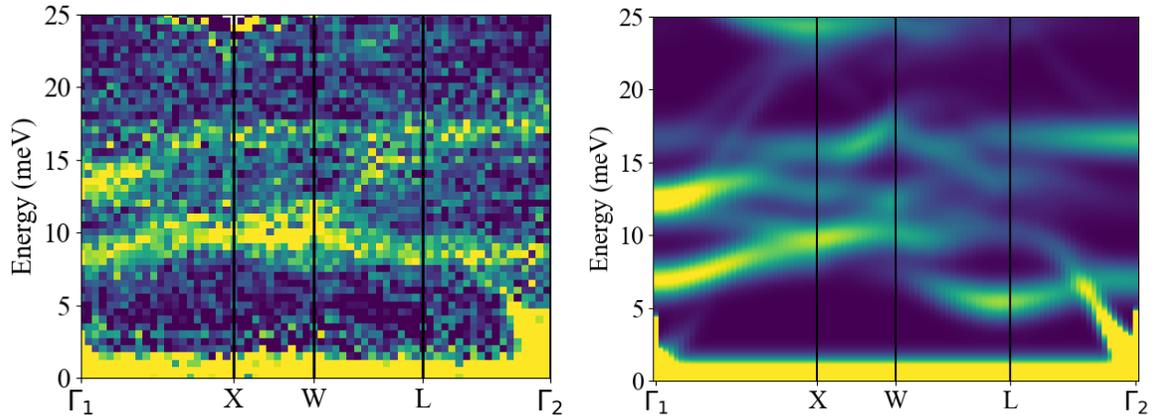

**Fig. 3.** The phonon dispersion along high symmetry directions. (a) The experimental data at 300 K and (b) the calculated intensities. The path mapped out in (a) and (b) is (-5,7,5), (-5,7,6), (-5,6.5,6), (-4.5,6.5,5.5) and (-4,6,6).

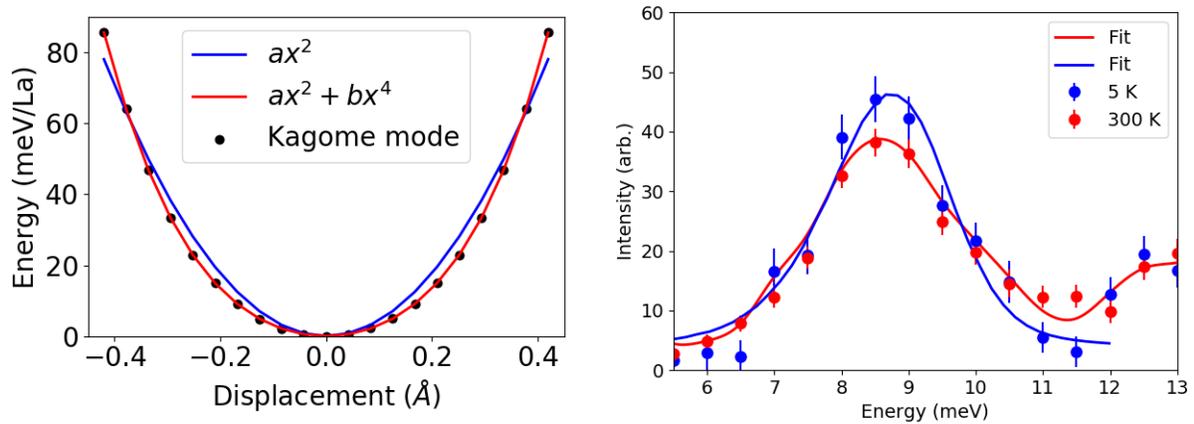

**Fig. 4.** Evidence for anharmonicty. (a) A frozen phonon calculation of the kagome mode at 6.87 meV. The normal mode displacement is 0.21 Å. A quadratic gives an extremely poor fit to the data while the introduction of a quartic term improves this dramatically. Here a=290.6(5) and b = 1104(3) while in the purely quadratic case a=442(10). (b) The low and high temperature kagome mode datasets and the corresponding fits.